\documentclass[12pt]{article}
\usepackage[dvips,dvipdf]{graphicx}
\usepackage{epsf}
\usepackage{color}
\usepackage{amssymb}
\usepackage{rotating}
\usepackage{psfig}
\usepackage{graphicx}

\oddsidemargin  -.cm
\evensidemargin-.cm
\newcommand{\beq}{\begin{eqnarray}}
\newcommand{\eeq}{\end{eqnarray}}
\newcommand{\be}{\begin{equation}}
\newcommand{\ee}{\end{equation}}

\begin{document}

\title{Influence of the ab-initio nd cross sections in the critical
heavy-water benchmarks}
\author{B. Morillon$^{a}$, R. Lazauskas$^{b}$, J. Carbonell$^{c}$ \\
{\small \emph{$^a$ CEA, DAM, DIF, F-91297 Arpajon, France}} \\
{\small \emph{$^b$ IPHC, IN2P3-CNRS/Universit\'e Louis Pasteur BP 28, F-67037 Strasbourg Cedex 2, France}} \\
{\small \emph{$^c$ CEA-Saclay, IRFU/SPhN, F-91191 Gif-sur-Yvette, France}} }
\maketitle

\begin{abstract}
The n-d elastic and breakup cross sections are computed by solving the three-body Faddeev
equations for realistic and semi-realistic Nucleon-Nucleon potentials. These
cross sections are inserted in  the Monte Carlo  simulation of the nuclear processes considered in the
International Handbook of Evaluated Criticality Safety Benchmark Experiments (ICSBEP). 
The  results obtained using thes {\it ab initio} n-d cross sections are compared with those  provided by the most renown international libraries.
\end{abstract}

{}

\section{Introduction}

\label{intr}

Monte Carlo simulations of neutron transport are broadly used in many domains of applicative nuclear physics. 
A key ingredient in these simulations is the knowledge of the nuclear cross sections. They are  determined by
evaluating the existing experimental data and compiled in the widely used nuclear data libraries. 
This is of course the most canonical and legitimate approach. However accurate experimental data
is not always available and for some particular reactions may be very costly or even impossible to attain.
In such a case, the models based on nuclear data evaluation and extrapolation
would lack in  credibility. On the other hand, and from a scientific point of view, the most satisfying
solution would be to get predictions of the desired cross sections by a reliable
theory derived from the first principles (i.e. nucleon-nucleon interaction).
Such idealistic approach still remains somehow utopic both due to the longstanding
puzzle of the nucleon-nucleon interaction and to our inability to solve the underlying
scattering problem for complex nuclear systems.

Nevertheless during the last few decades reliable theoretical tools have
been developed to describe few-nucleon scattering starting from the
nucleon-nucleon interaction. In particular neutron-deuteron system is well
explored, enabling accurate and complete description both for the elastic
and three nucleon break-up processes. Furthermore neutron-deuteron cross
sections turn to be predetermined by the well controlled long range part of
the nucleon-nucleon interaction. This fact makes us believe that the n-d cross
sections obtained from the first principles can be successfully applied in
Monte Carlo simulations as well as  to enrich the evaluated nuclear data libraries.

The paper is organized as follows. 
We present in Section 2 the theoretical tools
used to solve the three-body problem and to computed the corrresponding cross sections.
The results  provided  by some selected nucleon-nucleon potentials
are compared to the existing experimental data in Section 3.
Section 4 is devoted to describe the different sets of heavy water benchmarks for which
the Monte Carlo simulations will been performed.
The results of the multiplication factor $K_{eff}$  obtained by  Monte Carlo simulations including the ab initio cross sections
are presented in section 5. Some  conclusive remarks are finally drawn in Section 6.


\section{Computing the nd cross sections}


Our aim in this section is to describe the theoretical tools we have used to
compute the \textit{ab initio} neutron scattering on a deuteron, in the non
relativistic quantum mechanical framework provided by the Faddeev equations~%
\cite{Fad_60}.

\subsection{Kinematical variables}

We consider an incoming neutron ($m$) with laboratory kinetic energy $T_L$
and momentum $p_L=\sqrt{2T_Lm}$ impinging on a deuteron considered at rest. 
We are interested in the energy domain of $T_L\in[0,30]$ MeV and restrict to
the non-relativistic kinematics.

In order to simplify the dynamical equations, we have considered that  proton and neutron
have identical mass $m_p=m_n=m$, and take ${\frac{\hbar^2}{m}}=41.47$
MeV fm$^2$. For the neutron energies exceeding $T_L\geq\frac{3}{2} B_d=3.34$
MeV, where $B_d$ is the deuteron binding energy, the three particle break-up
channel is open, whereas at lower energies only elastic scattering of neutron is possible. 

\begin{figure}[h!]
\vspace{-.5cm}
\begin{center}
\epsfxsize=5.cm\epsfysize=5.cm\mbox{\epsffile{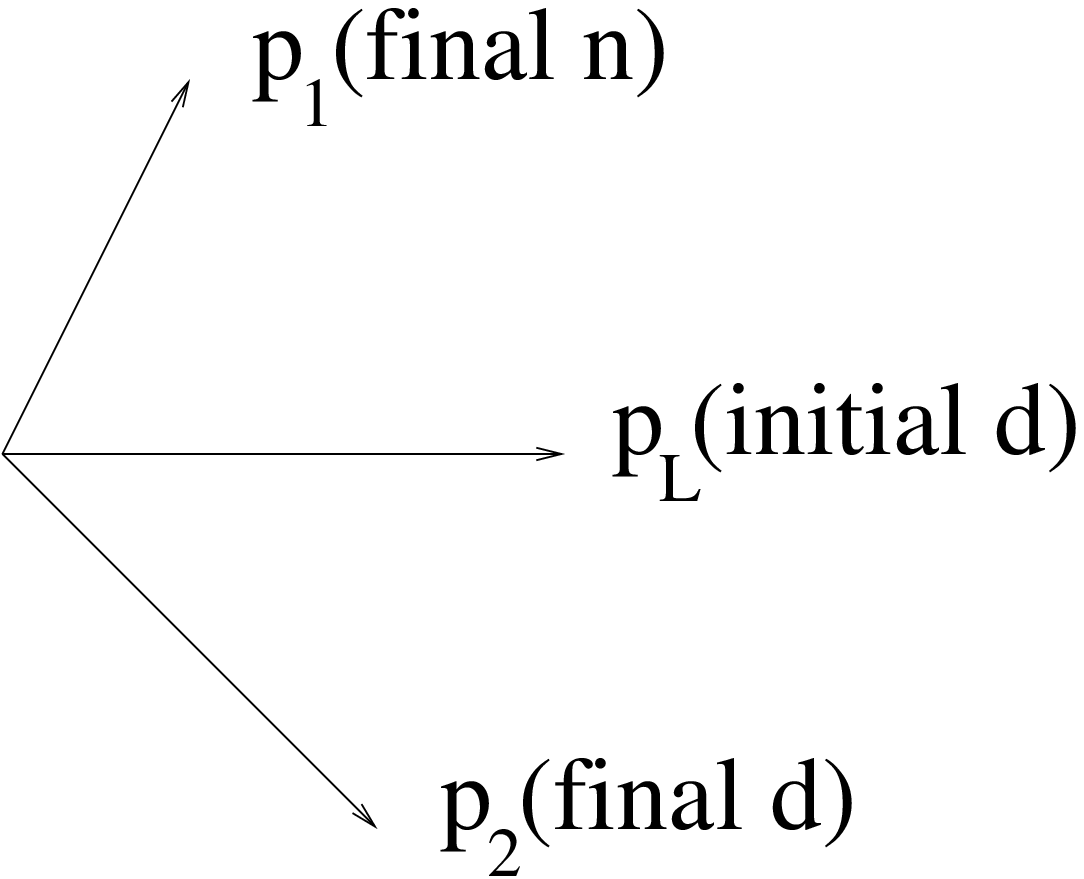}} \hspace{2.5cm}
\epsfxsize=5.cm\epsfysize=5.cm\mbox{\epsffile{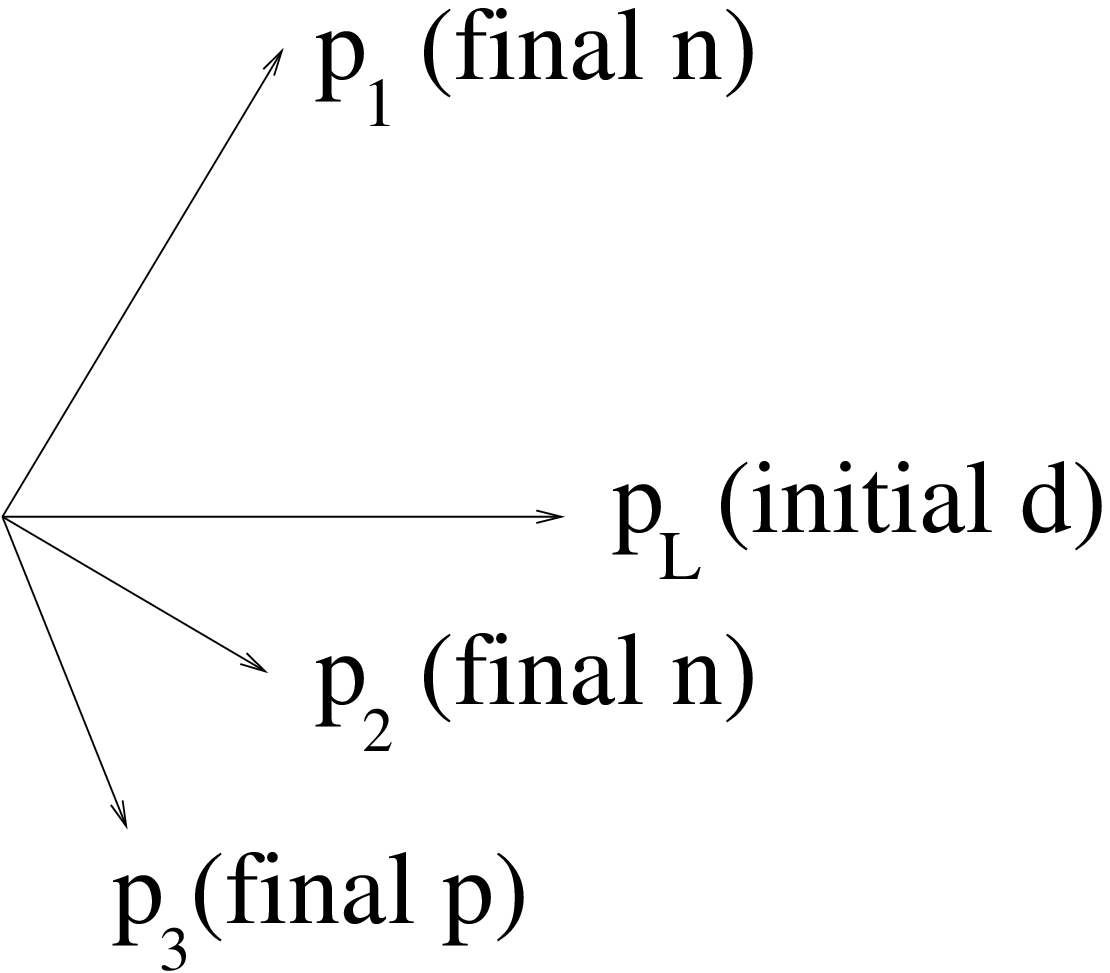}}
\end{center}
\caption{Kinematical laboratory variables in the n-d elastic scattering (left) and break-up reaction (right).}
\label{Kinematics}
\end{figure}

The momentum conservation laws for the elastic and break-up processes in the laboratory frame  are  displayed in Fig.  \ref{Kinematics}.

In the elastic channel the final state is determined by the momenta of the
scattered neutron $\vec{p}_{1}$ and deuteron $\vec{p}_{2}$. These are a
priori six unknowns, satisfying the four dynamical equations 
\begin{eqnarray}
\vec{p}_L &=& \vec{p}_{1}+\vec{p}_{2} \cr {p^2_L\over2m}&=& {\frac{{p}^2_1}{%
2m}}+{\frac{p^2_2}{4m}}  \label{Etot_2}
\end{eqnarray}
letting two free kinematical variables

In the break-up channel, the final state is determined by the momenta of the
three outgoing particles. We will denote by $\vec{p}_{1}$ and $\vec{p}_{2}$
the neutron momenta (here considered as distinguishable particles) and by 
$\vec{p}_3$ the proton one. This gives a priori nine unknowns, constrained by the conservation laws 
\begin{eqnarray}
\vec{p}_L &=& \vec{p}_{1}+\vec{p}_{2} +\vec{p}_3 \\
{\frac{p^2_L}{2m}}-B_d&=& {\frac{p^2_1}{2m}}+{\frac{p^2_2}{2m}}+{\frac{p^2_3%
}{2m}}  \label{Etot_3}
\end{eqnarray}
and giving five free kinematical variables.

\bigskip There exists several possible ways to choose the independent
kinematical variables among the momenta $\vec{p}_{i}$ defining the final
state. It is useful to express all of them in terms of the relative Jacobi
momenta $\vec{p}$ and $\vec{q}$ defined by 
\begin{equation}
\begin{array}{lcl}
\vec{p} & = & {\frac{1}{2}}\left( \vec{p}_{2}-\vec{p}_{3}\right) \cr\vec{q}
& = & {\frac{1}{\sqrt{3}}}\left( {\frac{\vec{p}_{2}+\vec{p}_{3}}{2}}-\vec{p}_{1}\right) 
\end{array}
\label{pq}
\end{equation}%
and the direction $\hat{n}$ of the incoming neutron in the laboratory frame $%
\vec{p}_{L}=p_{L}\hat{n}$. It is worth noticing that the Jacobi momenta $%
\vec{p}$ and $\vec{q}$ thus defined are invariants in any Galilean
transformation of coordinates and take consequently the same value when
computed in the laboratory (RL) or in the center of mass (RCM) reference
frame.

\bigskip To express the break-up cross section we will chose the incoming
energy $T_L$ and the four scalar variables: 
\begin{equation}  \label{5v1}
T_{L},\qquad \Theta_b=\arctan\left({\frac{q}{p}}\right),\qquad \hat{n}\cdot%
\hat{p},\qquad\hat{n}\cdot\hat{q},\qquad\hat{p}\cdot\hat{q}
\end{equation}
The quantity $\Theta_b$ is called the break-up angle and measure the
relative energy distribution between the fragments.

In the elastic channel, the quantity ${\frac{\vec{p}_2+\vec{p}_3}{2}}$ can
be identified with the scattered deuteron momenta, $\vec{p}$ becomes an
internal variable and the corresponding cross section -- which can be
considered as a particular case of the break-up case -- is characterized by
the two kinematical variables 
\begin{equation}  \label{2v1}
T_{L},\qquad \hat{n}\cdot\hat{q}
\end{equation}

We will describe in what follow the numerical method we have used to compute
both  the elastic and break-up cross sections, having as the only input   a given nucleon-nucleon interaction.

\subsection{The formalism}

The n-d reaction is considered as a three particle problem for nucleons
interacting via pairwise potentials $\hat{V}_i\equiv V_i( \mid \vec{r}_j-%
\vec{r}_k\mid)$ and is solved by means of the Faddeev equations \cite{Fad_63}. They
consist on a set of coupled equations for the so called Faddeev amplitudes $%
|\Psi_i>$ 
\begin{eqnarray}
(E-\hat{H}_0-\hat{V}_1) |\Psi_1> &=& \hat{V}_1 ( |\Psi_2> + |\Psi_3>) \cr (E-%
\hat{H}_0-\hat{V}_2) |\Psi_2> &=& \hat{V}_2 ( |\Psi_3> + |\Psi_1>)
\label{FE} \\
(E-\hat{H}_0-\hat{V}_3) |\Psi_3> &=& \hat{V}_3 ( |\Psi_1> + |\Psi_2>) 
\nonumber
\end{eqnarray}
where $\hat{H}_0$ denotes the intrinsic three-body free Hamiltonian These
equations are equivalent to the Schr\"{o}dinger one but allows a proper
treatment of the 3-body scattering problem. The total three-body
wavefunction $\Phi$ is given by 
\[
|\Phi> = |\Psi_1> + |\Psi_2> + |\Psi_3> 
\]

In the case of identical particles, the three Faddeev amplitudes are related
to each other by the permutation operators $P^{\pm}$. The system (\ref{FE})
is formulated in terms of one of the amplitudes, say $\Psi\equiv\Psi_1$, and
reduces to a single equation 
\begin{equation}  \label{FADD3}
\left[E-H_0-\hat{V}\right] |\Psi> =\hat{V}(P^{+} + P^{-}) |\Psi>
\end{equation}
with $\hat{V}\equiv\hat{V}_1$ the unique interaction potential and the
wavefunction obtained by 
\[
|\Phi> =( 1 + P^{+} + P^{-} )|\Psi> 
\]

Equation (\ref{FADD3}) can be solved in configuration space by projecting it
on the set of Jacobi coordinates associated with the amplitude $\Psi_1$ 
\begin{equation}
\begin{array}{lcl}
\vec{x} & = & \vec{r}_2-\vec{r}_3 \cr \vec{y} & = & {\frac{2}{\sqrt3}}\left( 
{\ {\frac{\vec{r}_2+\vec{r}_3}{2}} - \vec{r}_1} \right)%
\end{array}\label{xy}
\end{equation}

\begin{figure}[h!]
\begin{center}
\epsfxsize=3.5cm\centerline{\epsfbox{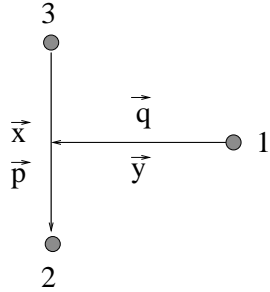}} 
\end{center}
\caption{Jacobi coordinates and momenta used in solving equation (\protect\ref{FADD4}). They are defined in  eqs. (\ref{pq})  and (\ref{xy}).} \label{Jacobi}
\end{figure}

The are the canonical conjugate of the Jacobi momenta introduced in the
 previous section -- eq. (\ref{pq}) --  and are displayed in Fig. \ref{Jacobi}.
 
It takes then the form of a partial differential equation on $R^6$ 
\begin{equation}  \label{FADD4}
\left[E-H_0-V(x)\right]\Psi(\vec x,\vec y) =V(x) (P^{+} + P^{-}) \Psi(\vec
x,\vec y)
\end{equation}
where 
\[
H_0= {\frac{\hbar^2}{M}} \left( \Delta_{\vec{x}} + \Delta_{\vec{y}} \right) 
\]
and 
\[
\Psi(\vec x,\vec y)= <\vec x,\vec y\mid\Psi> 
\]
is the projected amplitude. Notice that a non local coupling is induced by
the permutation operators on the right hand side of (\ref{FADD4}).

In order to solve equation  (\ref{FADD4}), the following boundary conditions \cite{Merkuriev_71,MGL_76}  must be implemented
\begin{equation}
\Psi (\vec{x},\vec{y})=\phi _{d}(x)f_{e}(T_{L},\theta _{e})\frac{e^{i\hat{q}%
\cdot \hat{y}}}{r}+A_{b}(T_{L},\theta _{e},\theta _{1},\theta _{2},\Theta )%
\frac{e^{ik\rho }}{\rho ^{5/2}}
\end{equation}%
were $\phi _{d}(x)$ is the deuteron wavefunction and $f_{e}(T_{L},\theta
_{e})$ and $A_{b}(T_{L},\theta _{e},\theta _{1},\theta _{2},\Theta )$ are
respectively the n-d elastic and break-up amplitudes we are interest in.
They are extracted from the  solution of (\ref{FADD4}) in the asymptotic region.

This solution  has been obtained numerically by means of the standard methods used in the Few-Body problems. 
They are based on a spline expansion of the Faddeev amplitudes which tranform the set of partial differential equations into an homogeneous linear system  \cite{These_Rimas,3n_2005}. 

As discussed in the previous section, these amplitudes depend on the
incoming neutron laboratory energy $T_L$, and the four angles denoted by 
\begin{eqnarray}
\theta_e&=&\hat{n}\cdot \hat{q} \cr \theta_1&=&\hat{n}\cdot \hat{p} \cr %
\theta_2&=&\hat{p}\cdot \hat{q} \cr \Theta &=&\arctan{\frac{p}{q}}
\end{eqnarray}
where $\hat{n}$ is the direction of the incoming neutron.

The elastic cross section is given by 
\begin{equation}  \label{dsig_e}
d\sigma_e= \mid f_e(T_{L},\theta_e)\mid^2 {\frac{9}{4 m p_L}} \; d\vec{p}_1 d%
\vec{p}_2 \; \delta(\vec{p}_1+\vec{p}_2 -\vec{P}_i)\delta(E_f-E_i)
\end{equation}
where $\vec{P}_i$ is the total momentum in the initial state ($\vec{P}_i=%
\vec{p}_L$ in the laboratory frame), $E_i$ and $E_f$ are respectively the
total energies on the initial and final states appearing in (\ref{Etot_2}).

In the center of mass frame, $\vec{P}_i=0$ and one can show that 
\[
{\frac{9}{4 m p_L}} \;d\vec{p}_1 d\vec{p}_2 \; \delta(\vec{p}_1+\vec{p}_1 -%
\vec{P}_i)\delta(E_f-E_i)= d\Omega_{\hat{q}} 
\]
where $\Omega_{\hat{q}}$ is the solid angle around direction $\hat{q}$. Then
one gets the usual expression for the elastic differential cross section 
\[
{\frac{d\sigma_e}{d\Omega_{\hat{q}}}} =\sigma_e(T_{lab},\theta_e) 
\]
with $\sigma_e(T_{lab},\theta_e)\equiv \mid f_e(T_{lab},\theta_e)\mid^2$

The break-up cross section has the form 
\begin{equation}  \label{dsig_b}
d^9\sigma_b={\frac{1}{8\pi}}\;{\frac{\hbar^2}{m}}\;{\frac{K_{3}}{k_{cm}^3}}
\;\mid A_b\mid^2\; d\vec{p}_1d\vec{p}_2d\vec{p}_3\;\delta(\vec{p}_L-\vec{p}%
_1-\vec{p}_2-\vec{p}_3)\delta(E_f-E_i)
\end{equation}
where momentum $K_{3}$ is related to the total energy $E_{3}$ of the three
nucleon system in the center of mass frame 
\[E_{3}={\frac{K_{3}^2}{m}} \]
$k_{cm}$ is related to the momentum $\vec{q}$ by $k_{cm}={\frac{2}{\sqrt3}}q$
and to the center of mass kinetic energy by 
\[
T_{cm}={\frac{\hbar^2}{m}} q^2 ={\frac{3}{4}}{\frac{\hbar^2}{m}} k_{cm}^2 
\]
and $E_i(E_f)$ are the total energies on the initial(final) states in (\ref%
{Etot_3}).

\section{Comparison with experimental results}\label{Comparison_with_experimental_results}

The only input in our calculations is the NN potential. We have considered three different interaction models: the so-called semi-realistic (MTI-III)
which has no any theoretical ground but give a good description of low
energy NN data and two  realistic ones (AV18 and INOY) which are, at
least partially, based on a one-boson exchange theory and describe the NN with a high degree of accuracy.

The semi-realistic MT I-III interaction \cite{MT13} is given by a simple sum
of two Yukawa potentials,  restricted to S-waves only but accounting for the
spin-dependent term. Despite his bare simplicity this interaction reproduces
reasonably well the bound states and low energy scattering observables of the A=2,
A=3 and A=4 nuclei, even without three-nucleon forces  \cite{PFG_PRC26_82,HG_NPA548_92,CC_PRC58_98}. It however
fails in describing processes where higher partial waves are involved.

As realistic NN interactions we have used the Argonne AV18~\cite{AV18_PRC_95} and INOY~%
\cite{DBPP_PRC67_03} potentials. They have both a much more complex
spatial and operator structure and depend on a large number of parameters ($\sim 40$) which are
fitted to describe almost perfectly the two-nucleon scattering data. The
AV18 is a local potential containing the pion-exchange tail plus the full
ensemble of operators allowed by symmetry to describe the intermediate and
short ranges part of the interaction. However, in order to provided a good description of some $A=3$ and $A=4$ zero energy observables
(like e.g.  the nd $J=1/2$ cross sections)  the inclusion of three-nucleon forces  is required. The strength
and form of these three-body forces depend on the two-body model they supposed to complete; together with Argonne AV18 potentials we have considered the so called Urbana UIX model \cite{UIX}.
The INOY potential superposes a local
one-pion exchange tail with a -- purely phenomenological -- strongly
non-local structure inside some internuclear distance $R_{c}\approx 1.5$ fm and reproduces  well, alone, the overal low energy dynamics of 3- and 4-nucleon states.

\begin{figure}[h!]
\begin{center}
\mbox{\epsfxsize=18.cm\epsffile{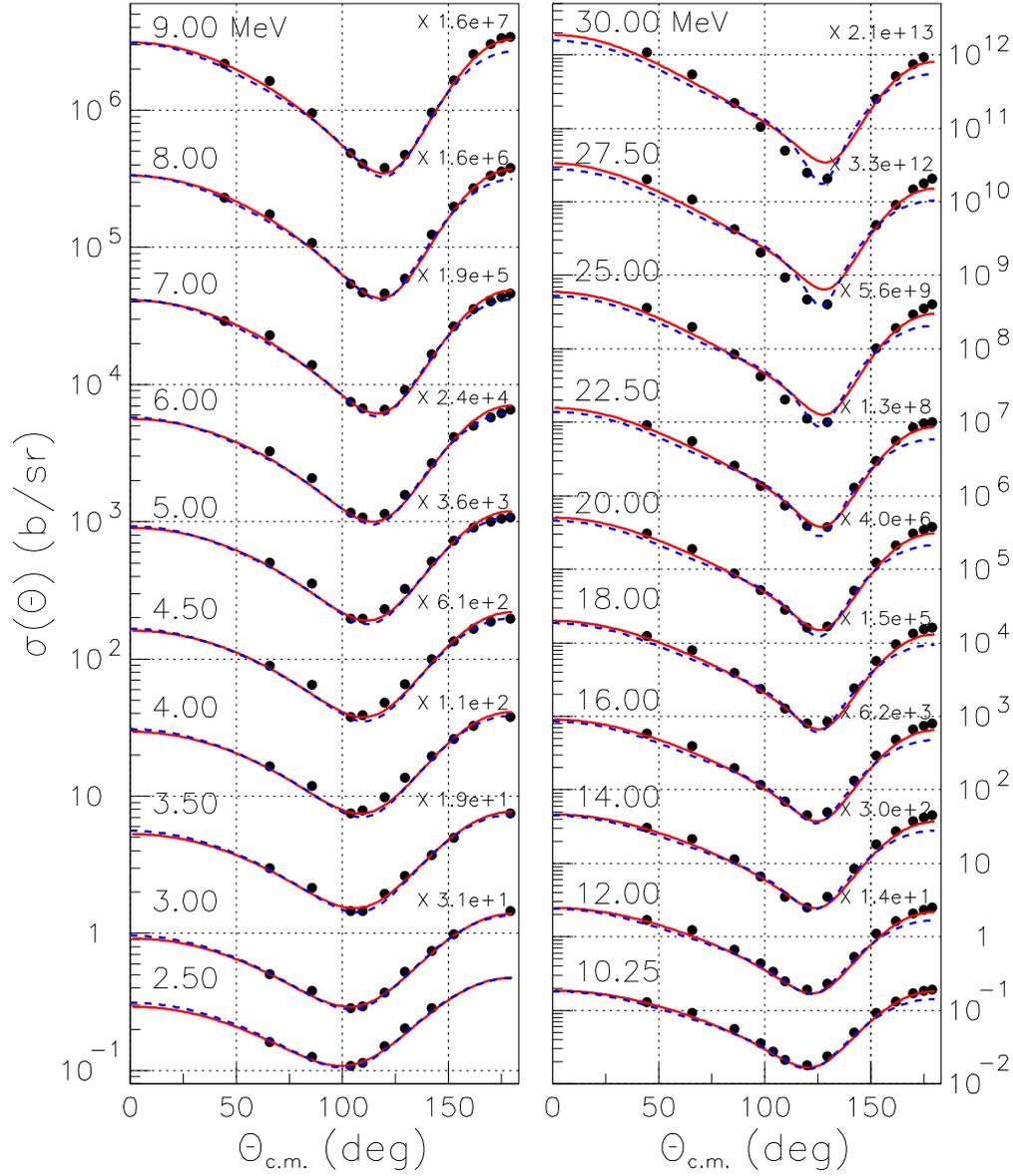}}
\caption{Elastic differential n-d  cross section using AV18 (solid red lines) and MT I-III NN (dashed blue lines) potentials is compared to experimental results taken from  \cite{SK_NPA398_83} } \label{fig_elakfk83}
\end{center}
\end{figure}

\begin{figure}[h!]
\begin{center}
\begin{center}
\mbox{\epsfxsize=14.cm\epsffile{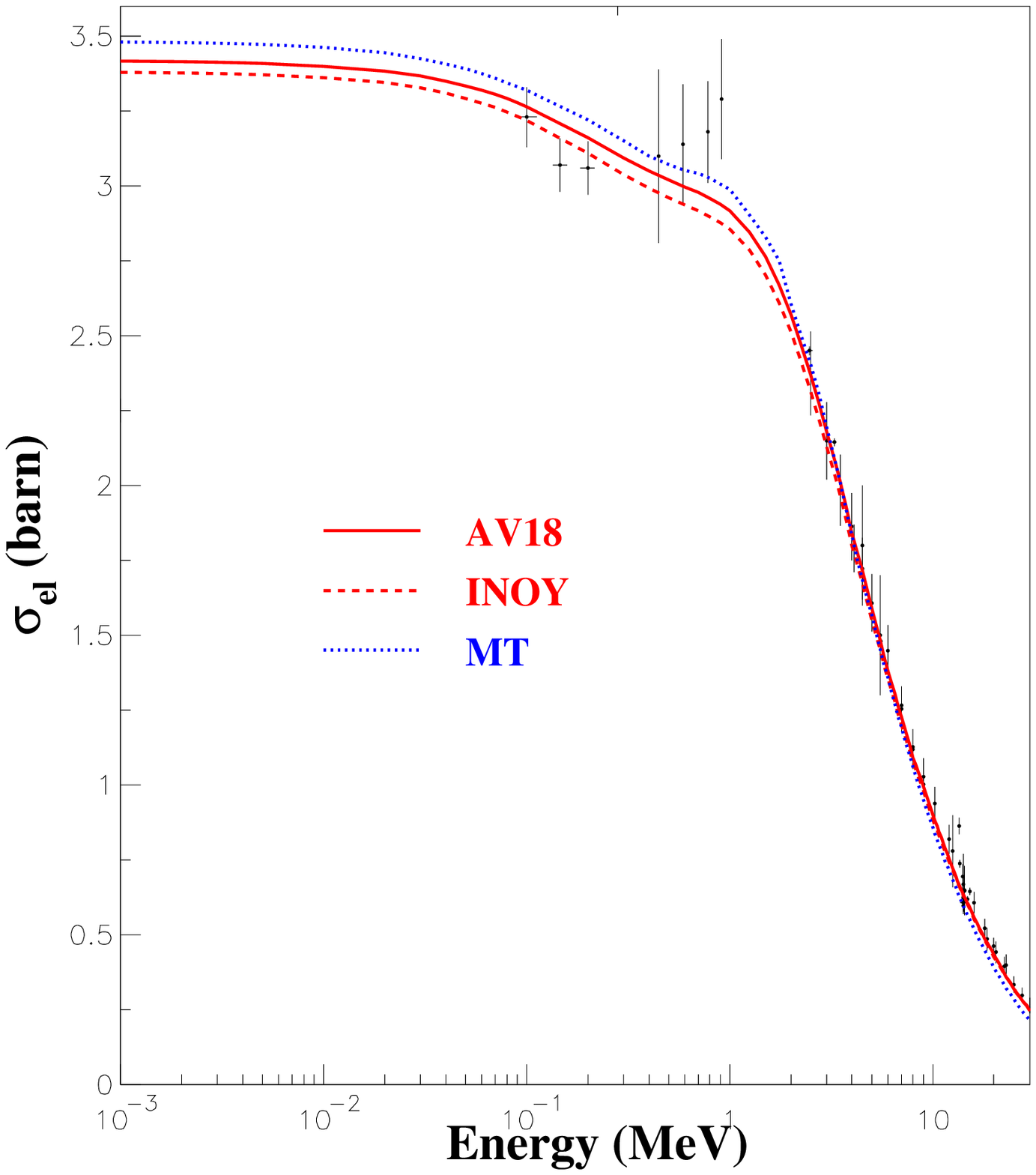}} \caption{Elastic n-d  integrated cross section using AV18, MT I-III and INOY NN potential}\label{fig_ela_av18mt}
\end{center}
\end{center}
\end{figure}

The results for the elastic differential and integrated cross sections are given respectively in Figs \ref{fig_elakfk83} and \ref{fig_ela_av18mt}. 
For the differential cross sections (Fig \ref{fig_elakfk83} ) only the AV18 (solide red line) and MT I-III (dashed blue lines) are displayed. The agreement 
of  AV18 with the experimental points is almost perfect and the differences with the simple MTI-III model are very small, visible only in the backward cross section.
The  integrated  elastic cross sections are given in Fig. \ref{fig_ela_av18mt}. At zero energy the three models differ each other by $\approx$2\% due to  
their differences in the nd scattering lenghts. These quantities have been calculated by several authors  \cite{CGM_FBS_93, Doleschal} and are summarized in Table \ref{tab:a0_nd} .
As one can see the quartet ($J=3/2$) value is one order of magnitude greater than the doublet ($J=1/2$) one and when computing the cross sections the quartet channel
 is furhtermore enhanced
by a larger statistical factor. The low enegy nd scattering is thus dominated by the  $J=3/2$ observales, which turn to be not sensible to the  three-nucleon forces. 
The MTI-III sliglhty overestimates (by 1\%)  the experimental
quartet scattering length while both AV18 and INOY are equal to each other and in full agreement with data. One can  also remark that, 
in absence of  UIX three-nucleon forces, the doublet value of AV18 sizeably overestimates the physical one.
These facts explain  the behaviour of the low energy cross sections displayed in Fig. \ref{fig_ela_av18mt}.
It is worth noticing however that the three considered models are all still compatible with the existing -- unfortunately not very accurate -- data for $E$ below 10 MeV.

\begin{table}[tbp]
\caption{Neutron-deuteron (nd)  scattering lengths (in fm) calculated using MT I-III, AV18 (+UIX) and Doleschall (INOY03) potentials.}\label{tab:a0_nd}%
\begin{center}
\begin{tabular}{lcc}
               & $^2a_{nd}$ (fm)& $^4a_{nd}$ (fm) \\ \hline
MT I-III   & 0.702          & 6.44            \\
INOY03       & 0.523      & 6.34            \\
AV18     & 1.26           & 6.34           \\
AV18+UIX & 0.595          & 6.34            \\
Exp.     & 0.65$\pm$0.04  & 6.35$\pm$0.02  \\
\end{tabular}
\end{center}
\end{table}

\begin{figure}[h!]
\begin{center}
\mbox{\epsfxsize12.cm\epsffile{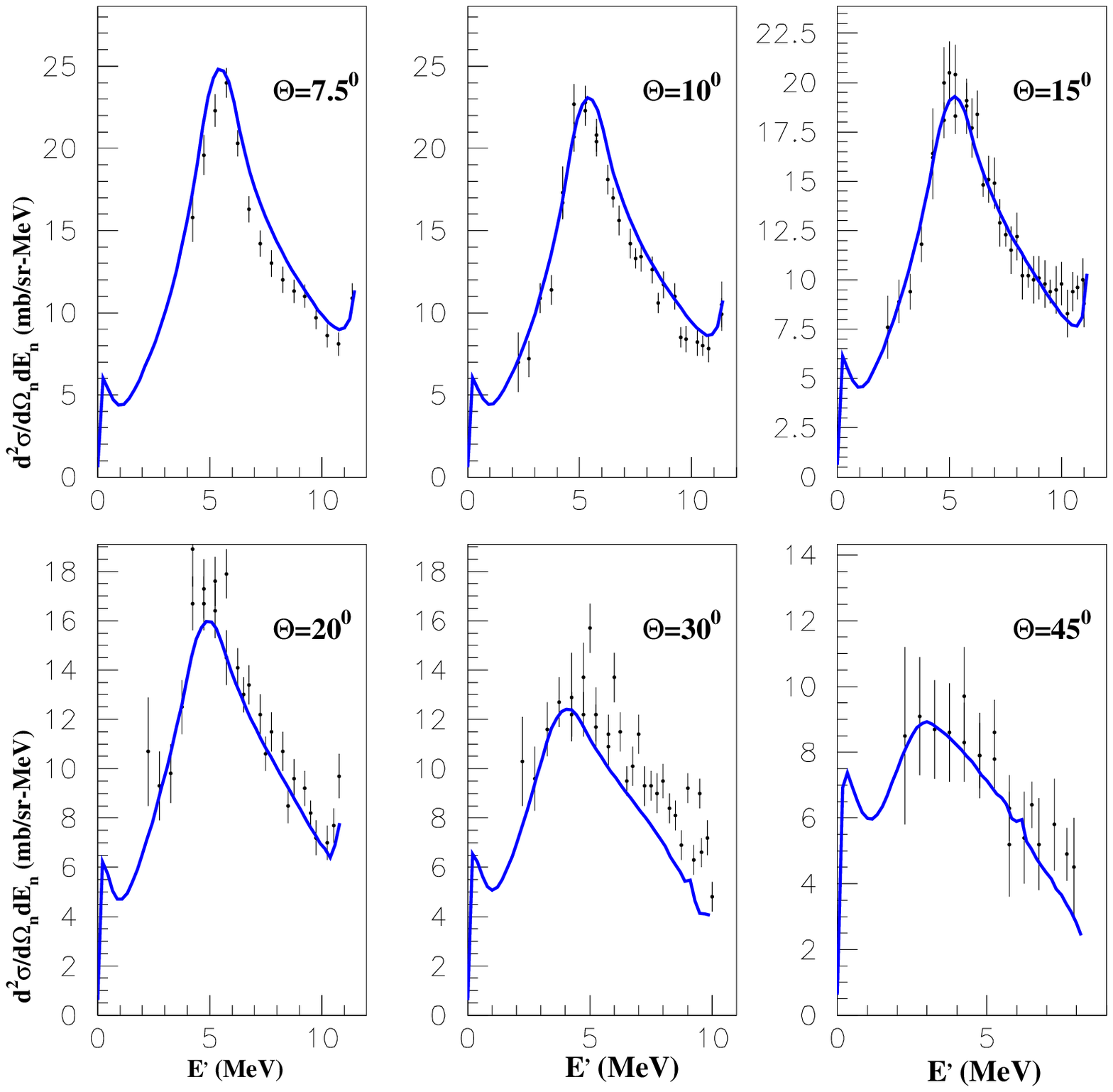}}
\caption{Neutron energy spectra ${d^3\sigma\over d\Omega dE}$  of the breakup reaction 
$^3H(n, 2n)H$  at neutron incident energy of 14.1 MeV and different angles  \cite{Zurich_NPA117_68} using MT I-III NN potential} \label{fig_daden2n14100kevnew}
\end{center}
\end{figure}

\begin{figure}[h!]
\begin{center}
\mbox{\epsfxsize12.cm\epsffile{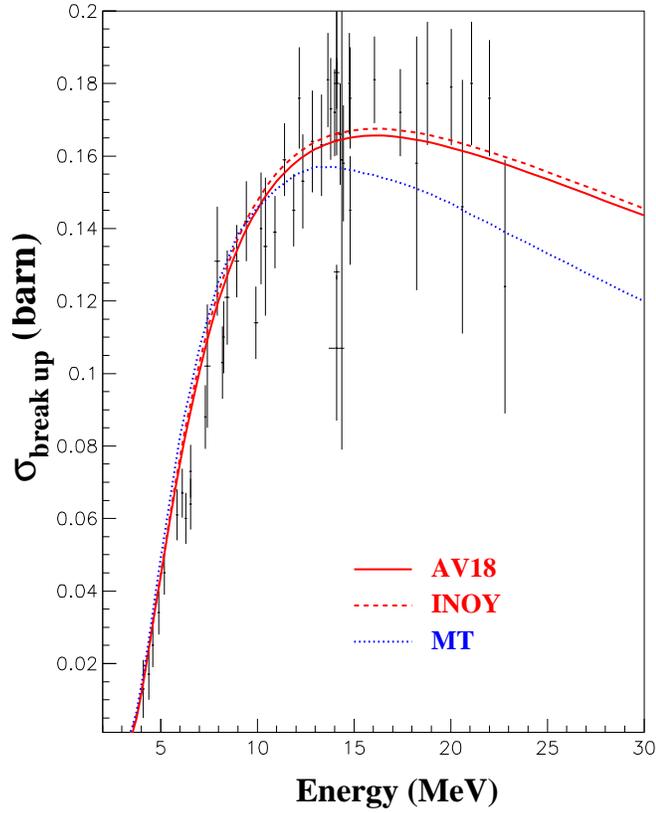}} \caption{Integrated nd break-up cross section using AV18, MT I-III and INOY NN potential}\label{fig_evalbup}
\end{center}
\end{figure}

\bigskip
The results of the differential and integrated break up cross sections are displayed respectively in Figs \ref{fig_daden2n14100kevnew} and \ref{fig_evalbup}. 
We have presented in Fig. \ref{fig_daden2n14100kevnew}  the neutron energy spectra ${d^3\sigma\over d\Omega dE}$  of the breakup reaction 
$^3H(n, 2n)H$  at neutron incident energy of 14.1 MeV and different angles \cite{Zurich_NPA117_68}.
As one can see the MTI-III model reproduces  well the data.
The integrated breakup cross section   as a function of the laboratory kinetic energy  displayed in Fig  \ref{fig_evalbup} is also well reproduced by all models up to $E\approx 15$ MeV
although beyond this energy, MTI-III   suffers from its lack of P-wave interaction.

\begin{figure}[h!]
\begin{center}
\begin{minipage}[h!]{7.2cm}
\begin{center}
\mbox{\epsfxsize=7.cm\epsffile{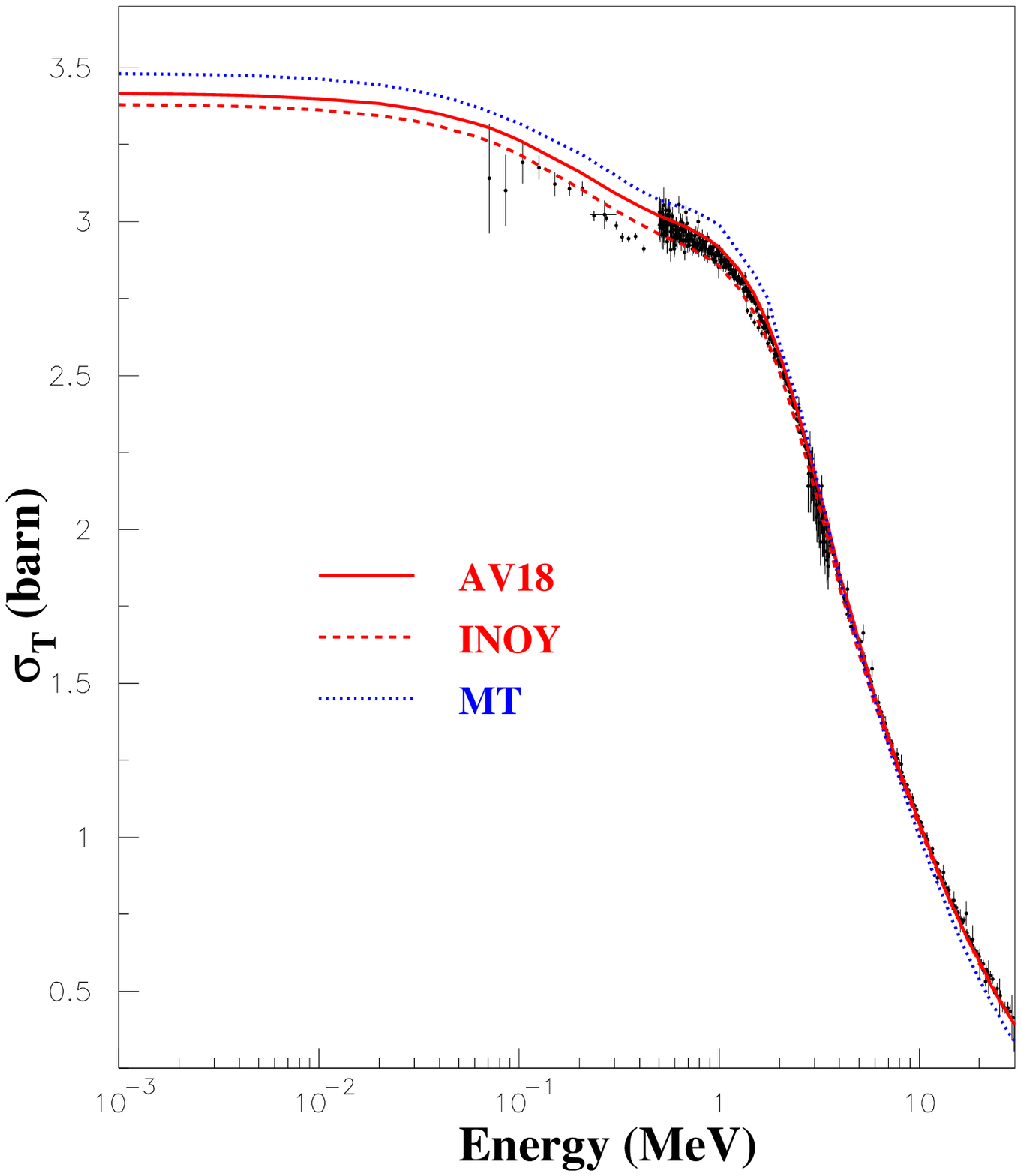}}
\caption{Total (elatic + breakup) n-d cross section using AV18, MT I-III and INOY NN potential}  \label{fig_tot_av18mt}
\end{center}
\end{minipage}
\hspace{0.2cm} 
\begin{minipage}[h!]{7.2cm}
\begin{center}
\mbox{\epsfxsize=7.cm\epsffile{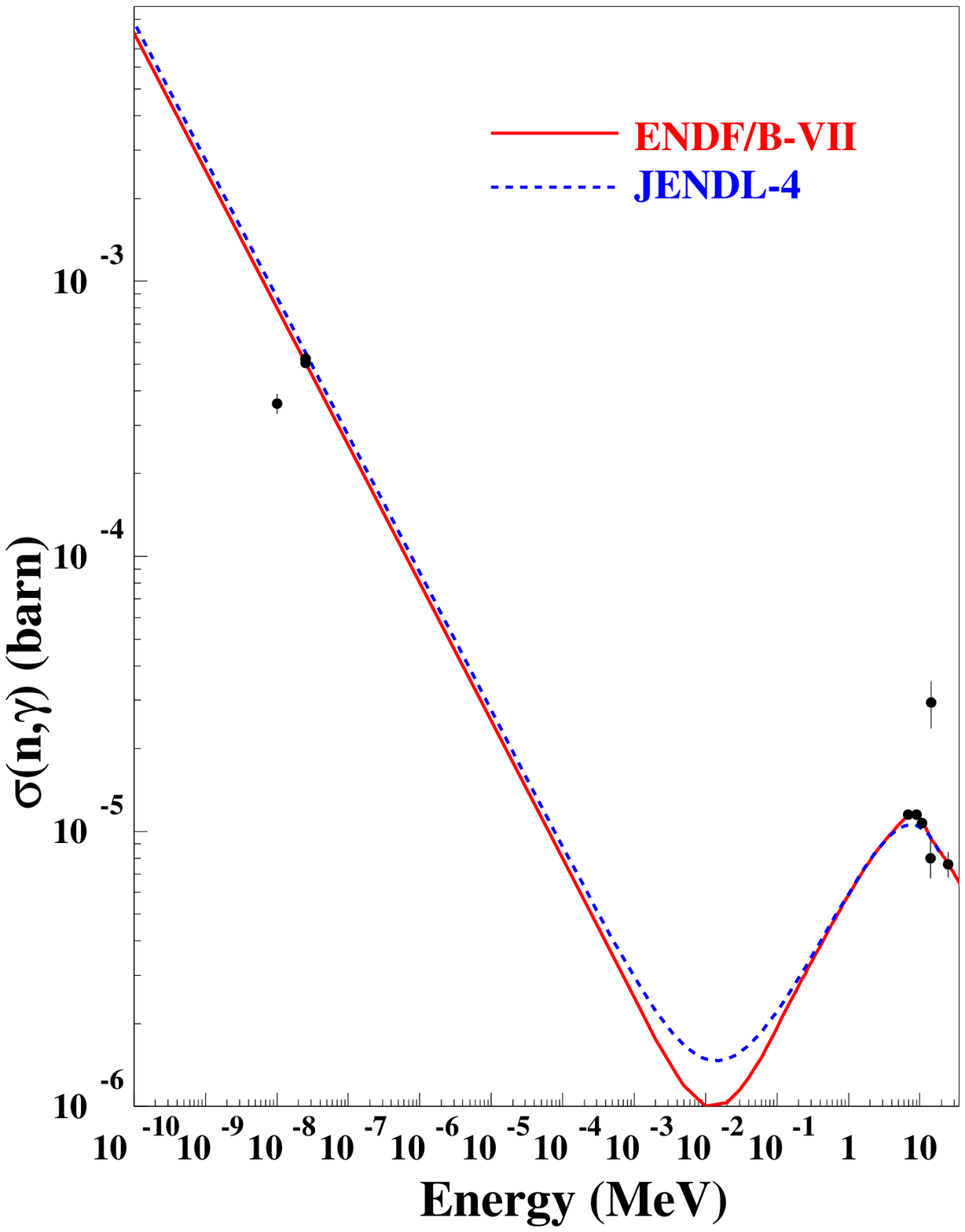}} \caption{Capture cross section from ENDF/B-VII and JENDL-4 evaluations}\label{fig_capture}
\end{center}
\end{minipage}
\end{center}
\end{figure}

The total (elastic plus break up) cross sections are given in Fig \ref{fig_tot_av18mt}. 
This quantity, of crucial importance in the Monte Carlo simulations, is however dominated by the elastic part
and do not deserve additional comments.

Among the deuteron cross sections, the only one that we have not
computed in this work is the neutron capture.  This cross section is however needed in the Monte Carlo  simulations described in next section. \color{black}
In order to fill this lack, we have taken this cross section from the ENDF/B-VII library~\cite{ENDFB70}. It is displayed in
Fig \ref{fig_capture} and -- in order to evualuate a systematic error -- it has been compared with the same results given by the JENDL-4 library~\cite{JENDL4}.

\bigskip
In summary, one can see that AV18 and INOY potentials are in perfect agreement  both for the elastic and for
break-up cross sections. They acurately reproduce the bulk of experimental data, except maybe for 
some particular break up geometries. This is a general result for any realistic
interaction model, which can accurately reproduce the two-nucleon scattering data. 
Indeed, due to the large deuteron size, the low energy n-d scattering cross section  is only
sensible to the well controlled long-range part of the two nucleon interaction
and is little affected by its short range off-energy shell structure, which
differs for realistic models. On the contrary the MT I-III potential
underestimates the experimental total cross sections starting from $E\approx 15$ MeV due to the absence of P-wave contribution.

\section{Heavy water benchmarks}\label{heavy_water_benchmarks} 

In this work we study the possibility of using the theoretically obtained n-d cross sections in the nuclear facilities simulations. 
The validity of the various nuclear data libraries as well as the
propagation codes are usually tested by trying to reproduce the results of the specially designed Benchmark Experiments. 
The heavy water benchmarks, being
strongly sensitive to the n-d scattering, constitute the testground we are looking for. 
We will  describe hereafter the technical details of the  benchmarks that we have tried to reproduce.

\subsection{Description of selected benchmarks}

The description of all the benchmarks we have considered in this work, is
taken from the International Handbook of Evaluated Criticality Safety Benchmark Experiments (ICSBEP Handbook)~\cite{ICSBEP08}. 
These benchmarks -- which concern only experiments
with deuterium --  are extracted from two different parts of this handbook:  from the Part II entitled "Highly Enriched Uranium (HEU)"
and from the Part IV denoted "Low Enriched Uranium systems (LEU)". 

Each of these two parts is in its turn separated into four sections, according to the physical form of the fissile material. 
For the deuterium experiments we are interested on, the fissile material can be
in form of metal (MET), compound (COMP) or solution (SOL). Moreover, each of
these four sections is subdivided into fast (FAST), intermediate (INTER), thermal
(THERM), and mixed (MIXED) spectra systems, depending on the neutron energy range at which the majority of the fissions occur.

In order to investigate  the influence of the  n-d cross section -- described in
previous section --  in the simulation codes,  we have considered the bechmarks proposed by the  ICSBEP Handbook: they consist in 45 experiments from Part II (HEU) and 35 from
Part IV (LEU). These experiments provide a consistent basis for evaluating the impact of the n-d cross sections on the reactivity. 
We will give in what follows,  a short description of each type of experiment. A more detailed analysis can be found in~\cite{ICSBEP08}.

\subsubsection{Highly enriched Uranium dioxide cylinders immersed in
mixtures of light and heavy water~(HCM002)}

In this series, 23 critical experiments were performed in 1985 at the Solution
Physics Facility of the Institute of Physics and Power Engineering (IPPE),
Obninsk, Russia. For the benchmark numbers 1, 4, 5, 8 and 9, the measurements were
made using 19 highly enriched uranium cylinders (with diameter 8.3 cm) immersed in
light water, whereas three different mixtures of light and heavy water were
used for the other cases. A thick bottom reflector of light water was used
in all the 23 critical experiments, while in fourteen of them an additional
lateral light-water reflector was employed. These experiments have been
originally denominated as HEU-COMP-MIXED-002 in~\cite{ICSBEP08}. The
abbreviation HCM002 is used in our figures.

\subsubsection{Cylindrical experiments using HEU plates reflected by lithium
deuteride~(HMF063)}

The Comet Universal Critical Assembly Machine was used to perform a series
of HEU critical experiments in 1955 at Los Alamos Scientific Laboratory. The
cylindrical assemblies were composed of plates of HEU metal, with 2.54~cm
thick reflector of $^6LiD$ and $LiD$. These experiments have been originally
denominated as HEU-MET-FAST-063 in~\cite{ICSBEP08} and are denoted by HMF063 in our figures.

\subsubsection{Reflected uranyl-fluoride solutions in heavy water~(HST004)}

They consist of six experiments in which heavy-water reflecting spheres of
uranyl-fluoride have been used. The atomic ratio of deuterium to $^{235}U$
ranged from 34 to 430. The denomination used in~\cite{ICSBEP08} is
HEU-SOL-THERM-004 and has been abbreviated  as HST004 in our figures.

\subsubsection{Unreflected cylinders of Uranyl-Fluoride solutions in heavy water~(HST020)}

In this benchmark, five assemblies of bare cylinders containing a ratio of deuterium
to $^{235}U$  varying from 230 to 2080 were considered. 
The denomination used in~\cite{ICSBEP08} is HEU-SOL-THERM-020 and corresponds to HST020  in our figures .

\subsubsection{RB reactor (HCT017, LMT001, LMT002, LMT015)}

An uranium heavy water critical assembly, known as the RB reactor, has a core  made of metal uranium
fuel elements which are displayed in an aluminum cylindrical tank filled with heavy-water moderator and reflector. 

The core of this reactor may vary according to the structure and type of its constituent elements (natural or
enriched metal uranium, placed in square lattices with various pitches) and give rise to the following configurations:

\begin{itemize}
\item \textbf{Lattices of 80\%-enriched Uranium   in heavy water~(HCT017)}

Lattices of fuel elements -- having the form of two concentric cylinders of
146.9 cm height and inner/outer diameters equal 3.1~cm/3.5~cm respectively --
are used. The outer cylinder is filled with uranium enriched up to 80\%,
while the inner one is filled with heavy water when it is  immersed in the
container. The number of fuel elements may be varied. The experimental
sequence is denominated as HEU-COMP-THERM-017 in~\cite{ICSBEP08} and
abbreviated by HCT017 in our work.

\item \textbf{Natural-Uranium rods in heavy water~(LMT001)}

The reactor core is made up of 208 metallic natural-uranium rods placed in a square lattice pitch of 12~cm and immersed in a cylindrical tank of aluminum
filled by heavy-water. The denomination in~\cite{ICSBEP08} is LEU-MET-THERM-001 and is abbreviated as LMT001 by us.

\item \textbf{Lattices of 2\%-enriched Uranium   in heavy water~(LMT002)}

The same as~(HCT017) cases but the cylinder height may be varied and the uranium
is enriched up to 2\% only. It is denominated as LEU-MET-THERM-002 in~\cite{ICSBEP08} and abbreviated as LMT002 in this work.

\item \textbf{Fuel assemblies substituted in lattices of 2\%-enriched Uranium (LMT015)}

This core allows the use of several  2\%-enriched metallic uranium elements. 
The fuel rods have the same cylindrical shape but may have void cavity or may be filled with heavy-water. 
A total of 22 criticality experiments were carried out in three series with
different number of fuel elements and with or without void cavity.

Two series of 7 criticality experiments in each run were performed. The heavy water contained 1.50 \% (molar) fraction of light water. The
fuel elements were placed in a square lattice pitch of 8~cm for the first run and 16~cm for the second one.

The third run was made up of fuel rods placed in a square lattice pitch
of 16~cm and immersed in heavy water with 0.73 \% (molar) fraction of light water. 
The denomination in~\cite{ICSBEP08} is LEU-MET-THERM-015 and  is denoted by LMT015 in this study.
\end{itemize}

\section{Results of simulations}

In this section, we present the results of the  multiplication factor $K_{eff}$ 
computed using the nd cross sections described in Section \ref{Comparison_with_experimental_results}.
To this aim we have built three new evaluations (in ENDF-6 format) of the deuterium cross sections corresponding to 
the different  NN interactions that we have investigated: MTI-III, AV18, INOY.

The results of our evaluations for the selected benchmarks have been compared  with those given by the international libraries.

\subsection{Monte Carlo calculations}

The ICSBEP Handbook contains the input listings for each experiment to
implement the Monte Carlo simulation MCNP5~\cite{MCNP5} code in order to
calculate the multiplication factor $K_{eff}$. 

The input may be slightly modified to account for the various nuclei described by the
libraries. The nuclear cross sections, apart the nd one,  are taken from the ENDF/B-VII library.

The total computing time for the 90 selected experiments is approximately 16
hours on 48 processors to achieve a standard deviation less than 25 pcm for each benchmark.

\subsection{Thermal scattering laws $S(\protect\alpha ,\protect\beta )$}

Although this article is not devoted to discuss the thermal scattering laws,
commonly denoted $S(\alpha ,\beta )$, it is necessary to take them into
account in the simulations. Thermal effects have been simulated using the $S(\alpha,\beta )$
taken from the libraries  ENDF/B-VII and SAB2002~\cite{SAB2002}. 

The sensitivity study of the deuterium thermal scattering laws is presented in Fig.~\ref%
{fig_bench_sab}. The multiplication factors calculated using the SAB2002 data are slightly lower than those using ENDF/B-VII and
are in a slightly better agreement with the experimental results, especially for HCT017, LMT002 and LMT015 cases. 
This fact led us to chose the SAB2002 scattering laws  all along this study.

\begin{figure}[tbph]
\centerline{\psfig{figure=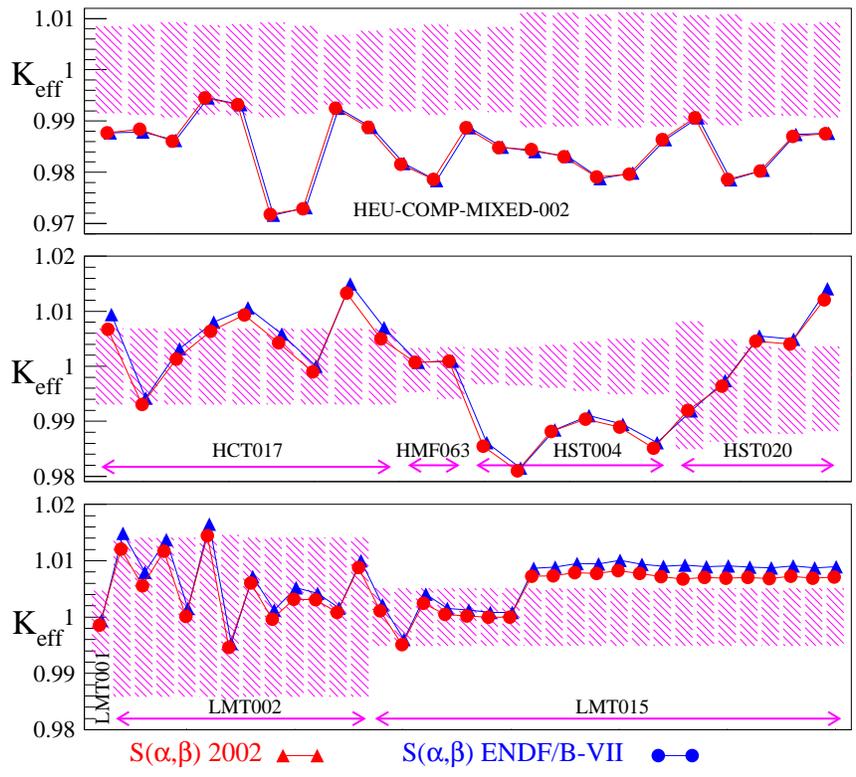,height=12.0cm}}
\caption{Comparison of criticality calculations when the thermal scattering laws $S(\protect\alpha,\protect\beta )$ are adapted from ENDF/B-VII~%
\protect\cite{ENDFB70} (red circles) and SAB2002~\protect\cite{SAB2002}
(blue triangles) libraries.}
\label{fig_bench_sab}
\end{figure}

\subsection{Deuterium evaluation from international libraries}

Many recent nuclear data libraries contain an evaluation of the deuterium
cross sections. The most recent ones are: the American ENDF/B-VII,
the Japanese JENDL-4 and the Chinese  CENDL-3.1 \cite{CENDL31}. 

\begin{figure}[h]
\begin{center}
\begin{minipage}[h!]{7.5cm}
\begin{center}
\mbox{\epsfxsize=7.5cm\epsffile{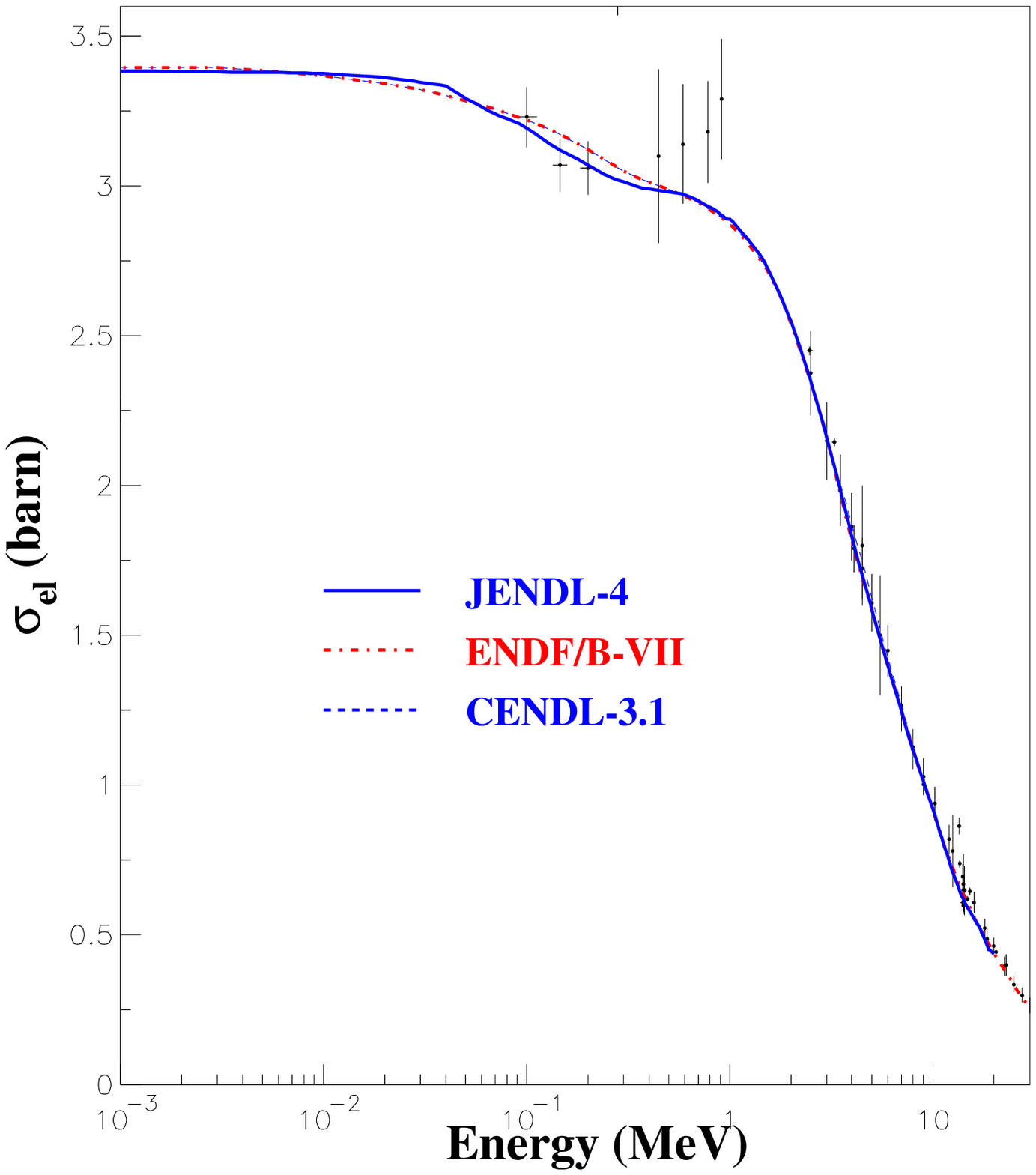}}
\caption{Elastic n-d  cross section taken from the libraries: ENDF/B-VII (red dashed-dotted line),
JENDL-4 (blue solid line) and CENDL-3.1 (blue dotted line).}  \label{fig_eval_ela}
\end{center}
\end{minipage}
\hspace{0.2cm} 
\begin{minipage}[h!]{7.5cm}
\begin{center}
\mbox{\epsfxsize=7.5cm\epsffile{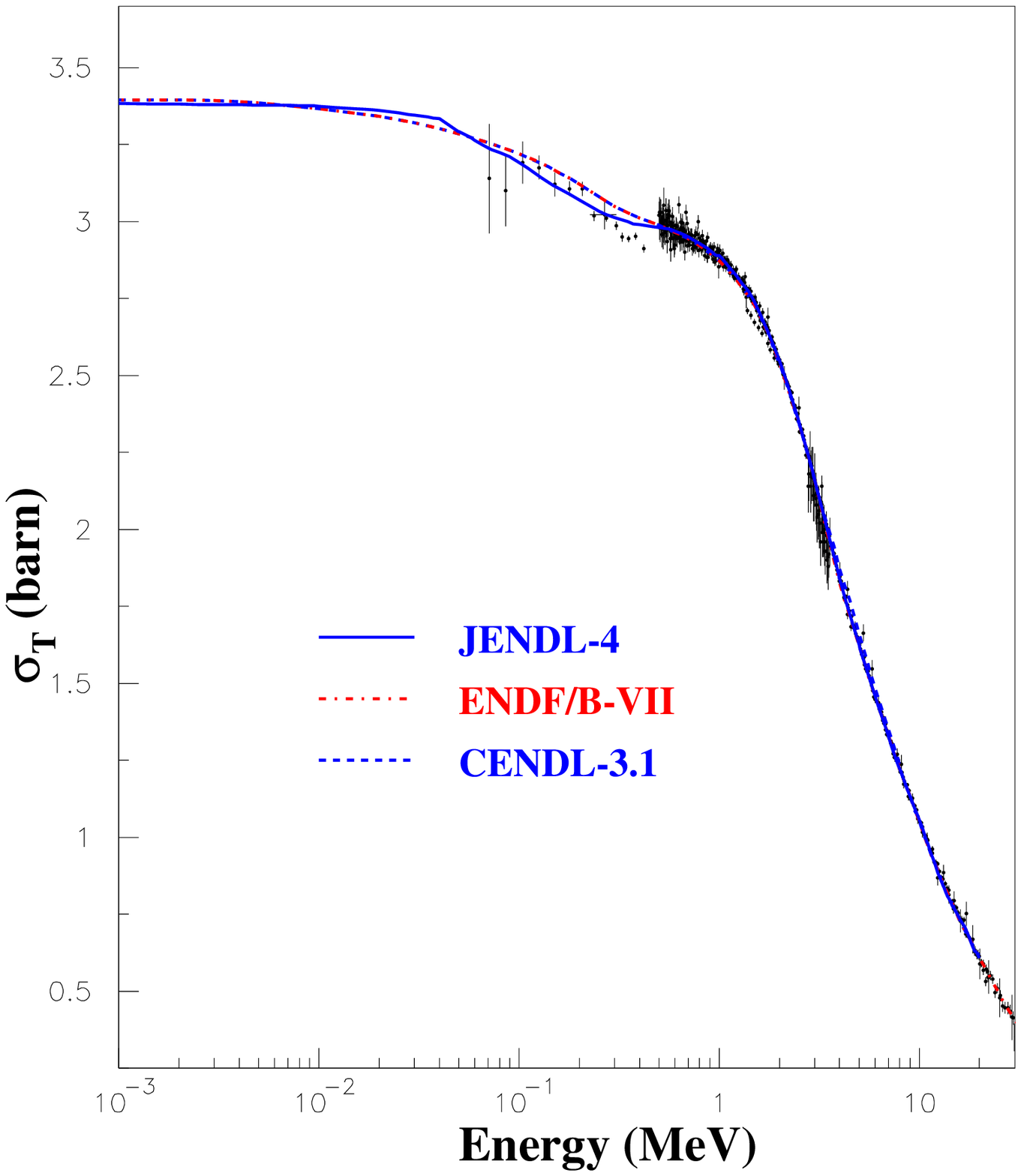}} \caption{Total n-d  cross section taken from the
libraries: ENDF/B-VII (red dashed-dotted line),
JENDL-4 (blue solid line) and CENDL-3.1 (blue dotted line).}\label{fig_eval_totale}
\end{center}
\end{minipage}
\end{center}
\end{figure}

These evaluations provide  a complete representation of the
nuclear data needed for simulating the neutron transport over an energy range which varies from $10^{-11}$ to 150 MeV in  the American library and from
$10^{-11}$ to 20 MeV  in the Japanese or Chinese ones.
The elastic and total n-d  cross
sections coming from these recent nuclear data libraries are given respectively
in Figs \ref{fig_eval_ela} and \ref{fig_eval_totale}.
The American and Chinese  evaluations of the  elastic and total n-d  cross sections
are identical in all the energy range and differ slightly from the Japanese one below 1 MeV, as can bee senn in these Figures.

\begin{figure}[tbph]
\centerline{\psfig{figure=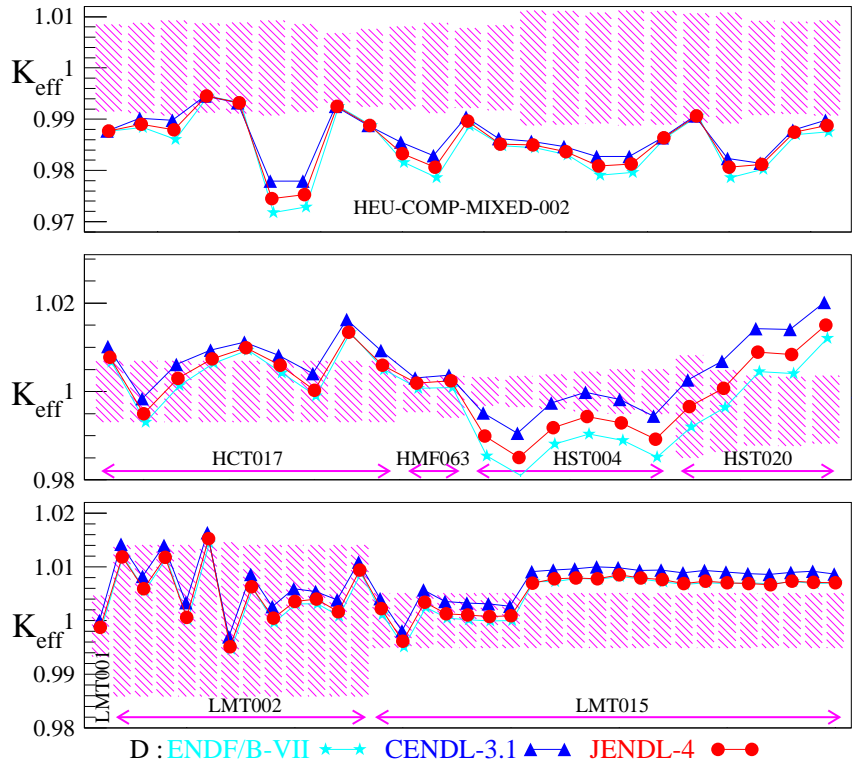,height=12.0cm}}
\caption{Sensitivity of the criticality calculation to the n-d cross
sections taken from the different libraries: ENDF/B-VII (cyan stars),
JENDL-4 (red circles) and CENDL-3.1 (blue triangles).}
\label{fig_bench_b7c3j4}
\end{figure}

In Fig.~\ref{fig_bench_b7c3j4} we compare the multiplication factors
calculated using the MCNP5 code with the deuterium cross sections taken
respectively from ENDF/B-VII (cyan stars), JENDL-4 (red circles) and
CENDL-3.1 (blue triangles); as indicated above all the other nuclear cross
sections are taken from the ENDF/B-VII library and $S(\alpha ,\beta )$
data is imported from SAB2002. In such a way, it is easier to compare the
different evaluations. The three deuterium evaluations lead to a
systematical underestimation of $K_{eff}$ values, from about 0.2 to 2.0 \%,
provided by the HCM002 benchmark measurement. 

The ENDF/B-VII evaluation of the deuterium cross sections provides systematically the lowest values whereas the
CENDL-3.1 evaluation provides the highest ones; the JENDL-4 result is in
between. 
The largest variation occurs for the HST004 and the HST020 benchmarks. 

\subsection{Our new evaluations of the deuterium cross sections}

\begin{figure}[tbph]
\centerline{\psfig{figure=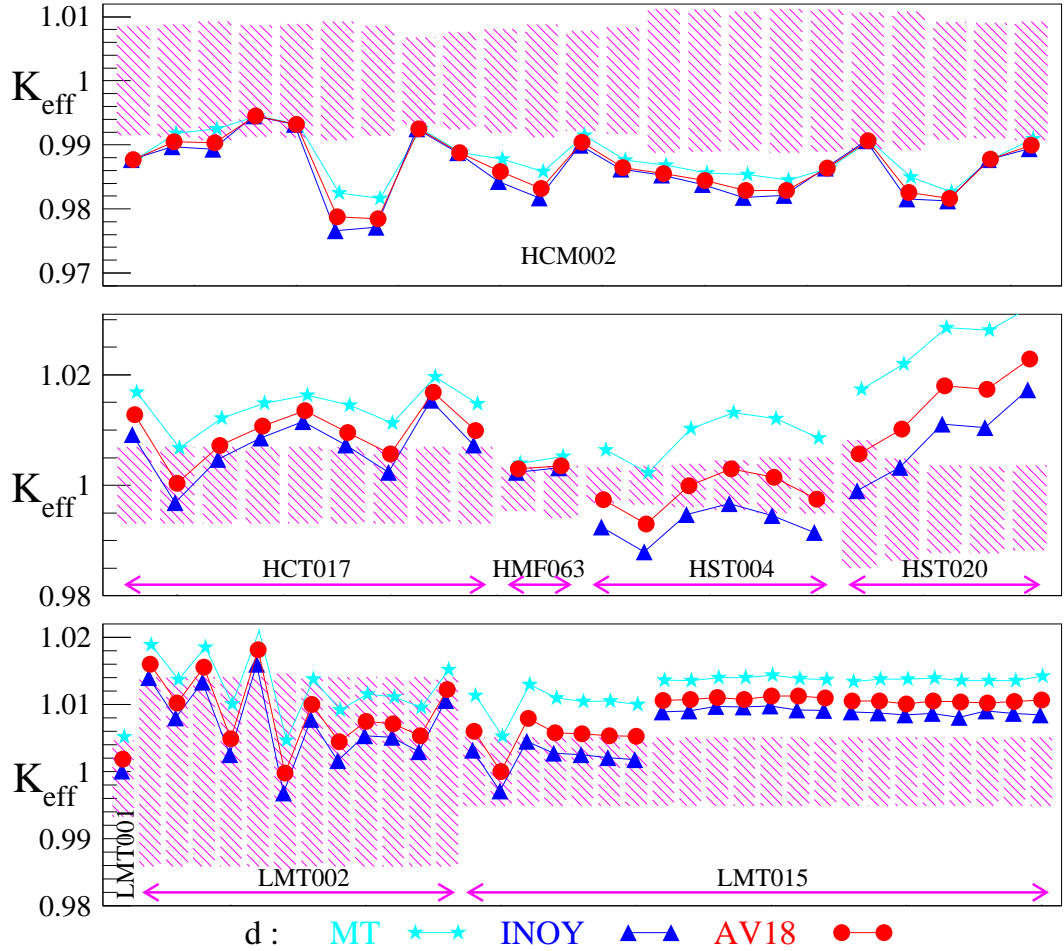,height=15.0cm}}
\caption{Multiplication factors $K_{eff}$  obtained in  Monte-Carlo simulations using the nd cross sections
given by different NN interaction models: MT I-III (cyan star symbols),
AV18 (red circles) and INOY (blue triangles). They are compared to experimental results of selected benchmarcks (hachured vertical bars).}
\label{fig_bench_b7dmtinoyav18}
\end{figure}

We have built a new deuterium evaluations inserting the cross sections  obtained by solving the Faddeev equations with the selected NN potentials. 
The $K_{eff}$ values corresponding to the different nucleon-nucleon interactions  are 
presented in Fig.~\ref{fig_bench_b7dmtinoyav18}. The MTI-III results are indicated by  cyan star symbols, AV18 by filled red circles and INOY by filled blue triangles.
They are  compared to the experimental results of the benchmarks (hachured vertical bars). 
Notice that for some particular configurations of the HCM002 series (e.g. colons 1,4,5,8,9  starting from left)  the three potentials gives identical
results. They correspond to the absence of deuterium in the corresponding  benchmark but they have been  nevertheless included in our simulations
as a consistency check.

As a general remark we see that the MT I-III calculations provide always higher reactivities, a direct consequence of its slightly overestimation of the low
energy nd cross sections (see Fig \ref{fig_ela_av18mt} ).
Except for the HCM002 series -- where the  disagreement  is slightly reduced --  
the MTI-III reactivity values  are above the experimental results for most of the examinated benchmarks.
Apart from this particular series, the agreement is systematically improved when  using realistic nucleon-nucleon interaction, such as AV18 or INOY.
The results displayed in  Fig.~\ref{fig_bench_b7dmtinoyav18} tend to favour INOY with respect to AV18 interaction, probably due
to its non requirement of three-nucleon forces (see results on Table \ref{tab:a0_nd}).
This fact demonstrates the importance of a realistic description of the n-d scattering.

We would like to notice however that none of the potentials is able to  describe  satisfactorily the whole set
of benchmarks. A systematic disagreement at the level of 1\% exists 
for some particular configurations of the HCM002, HST020 and LMT015 series which is difficult to attribute to a 
default of the interaction. The only possible explanation concerning the deuterium data
is to be search on the capture cross section or in the thermal scattering laws.
An inconsistency  of the benchmark themselves cannot be excluded. Is worth noticing for instance
that even in absence of deuterium (e.g. colon 1 from HCM002) the simulations cannot reproduce well   
the experimental result.

\begin{figure}[tbph]
\centerline{\psfig{figure=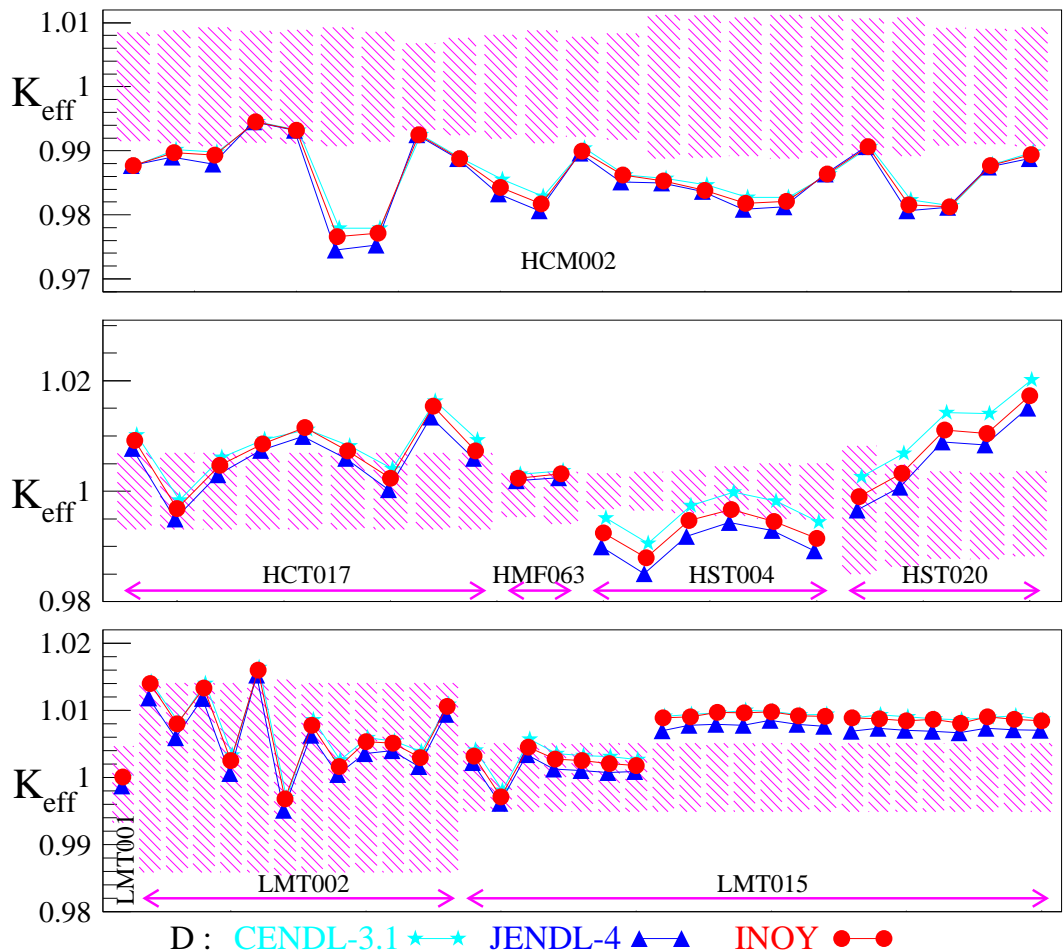,height=15.0cm}}
\caption{Sensitivity of the criticality calculation to the n-d  cross sections input: the {\it ab initio} INOY resuts (red circles) are compared
to those given by the  CENDL-3.1 (cyan stars) and JENDL-4 (blue triangles) libraries.} \label{fig_bench_b7dc3j4inoy}
\end{figure}

Finally we have compared in Fig.~\ref{fig_bench_b7dc3j4inoy} the  reactivity obtained using the INOY deuterium cross sections  (red circles)
with the calculations based on the international library cross sections CENDL-3.1 (cyan stars) and JENDL-4 (blue triangles). 
In general, the INOY results  settles  in between  those of these two libraries. 
We conclude from that, that the evaluations based on n-d cross sections obtained from
{\it ab initio} nuclear physics  calculations  are of the same quality than those provided by the best established international library.
It would be interesting to perform similar studies by computing other interesting  quantities  in the neutron transport simulations.

\section{Conclusion}

The aim of this paper was to check the validity of {\it ab initio} nuclear calculations
in the integral experiments involving critical benchmarks.

\bigskip
By solving the Faddeev equations in configuration space, we have computed the n-d differential elastic and break-up cross sections
using three different nucleon-nucleon interactions (MTI-III, AV18 and INOY).
These cross sections have been inserted in a Monte Carlo simulation of 
neutron transport to obtain the multiplication factor $K_{eff}$.  
The results have been compared to some selected experimental benchmark.

\bigskip
We have found that the semi-realistic MTI-III potential systematically overestimates
the reactivity and that the best agreement is provided by the non local nucleon-nucleon potential INOY.
These results fails however to   describe accurately the whole set
of benchmarks and a systematic discrecpancy (of the order of 1\%) exists 
for some particular configurations  which is difficult to attribute to a  default of the interaction. 

\bigskip
The  results obtained by the {\it ab initio}  INOY potential  have been compared with the calculations based on the international library cross sections
and are found to be of the same quality.
It would be interesting to perform similar studies by computing other quantities  which are relevant in the neutron transport simulations.

\section*{Acknowledgement}

This work was granted access to the HPC resources of IDRIS under the
allocation 2009-i2009056006 made by GENCI (Grand Equipement National de
Calcul Intensif). We thank the staff members of the IDRIS for their constant
help.



\begin{thebibliography}{99}
\bibitem{Fad_60} L.~D. Faddeev, JETP 39 (1960) 1459, Sov. Phys. JETP 12 (1961) 1014

\bibitem{Fad_63} L.~D. Faddeev, Mathematical Aspects of the Three-Body Problem in the Quantum Scattering Theory, Transl. by Israel Program for Scientific Translations Ltd. (1965), IPST Cat. No. 2158

\bibitem{Merkuriev_71} S.~P. Merkuriev, Theoretical and Mathematical Physics
8 (1971) 798 

\bibitem{MGL_76} S.~P. Merkuriev, C. Gignoux, A. Laverne, Ann. Phys 99
(1976) 30

\bibitem{These_Rimas} R. Lazauskas, PhD Thesis, Universit\'{e} Joseph
Fourier, Grenoble (2003);
http://tel.ccsd.cnrs.fr/documents/archives0/00/00/41/78/.

\bibitem{3n_2005} R. Lazauskas, J. Carbonell,   Phys. Rev. C71 (2005) 044004; nucl-th/0502037    


\bibitem{MT13} R. A. Malfliet and J. A. Tjon, Nucl. Phys. A127, 161 (1969)

\bibitem{PFG_PRC26_82} G. L. Payne, J. L. Friar, and B. F. Gibson, Phys.
Rev. C 26, 1385 (1982).

\bibitem{HG_NPA548_92} H. Kamada and W. Glo$\lnot$ckle, Nucl. Phys. A548,
205 (1992)

\bibitem{CC_PRC58_98} F. Ciesielski, J. Carbonell, Phys. Rev. \textbf{C58}
(1998) 58-74


\bibitem{AV18_PRC_95} R.B. Wiringa, V.G.J. Stoks, R. Schiavilla, Phys. Rev.
C 51 (1995) 38


\bibitem{DBPP_PRC67_03} P. Doleschall, I. Borb\'ely, Z. Papp, W. Plessas,
Phys. Rev. C 67 (2003) 064005

\bibitem{UIX} B. S. Pudliner, V. R. Pandharipande, J. Carlson, and R. B. Wiringa, Phys. Rev. Lett. 74, 4396 1995 .

\bibitem{CGM_FBS_93}
        J. Carbonell, C. Gignoux, S. P. Merkuriev,        Few-Body Systems, {\bf 15} (1993) 15-23

\bibitem{Doleschal}  R. Lazauskas, J. Carbonell, Phys. Rev C70 (2004) 044002; nucl-th/04080

\bibitem{Zurich_NPA117_68} M. BR†LLMANN, H . JUNG, D . MEIER und P . MARMIER, Nucl. Plys 117 (1968) 419 

\bibitem{SK_NPA398_83} P.~Schwarz, H.O. Klages, P.~Doll, B.~Haesner,
J.~Wilczynski, B.~Zeitnitz, and J.~Kecskemeti., Nucl. Phys. A398, 1 (1983)

\bibitem{ENDFB70} M.B. Chadwick \emph{et al.}, "ENDF/B-VII.0: Next
generation evaluated nuclear data library for nuclear science and
technology", Nucl. Data Sheets \textbf{107}(2006)2931.

\bibitem{JENDL4} K. Shibata \emph{et al.},"JENDL-4.0: A New Library for
Nuclear Science and Engineering", J. Nucl. Sci. Technol. \textbf{48}, 1-30
(2011).

\bibitem{ICSBEP08} International Handbook of Evaluated Criticality Safety
Benchmark Experiments, NEA/NSC/DOC(95)03, OECD Nuclear Energy Agency,
September 2008 Edition.

\bibitem{MCNP5} X5-MCNP-Team, ``MCNP - A General Monte Carlo NParticle
Transport Code, Version 5, Volume I: Overview and Theory,'\ Tech. Rep.
LA-UR-03-1987, Los Alamos National Lab, April 2003.

\bibitem{SAB2002} R. C. Little and R. E. MacFarlane, \textquotedblleft
SAB2002---An $S(\alpha ,\beta )$ Library for MCNP,\textquotedblright\
X-5-03-21(U), Los Alamos National Laboratory (February 3, 2003).

\bibitem{CENDL31} Z.G. Ge \emph{et al.}, "The Updated Version of Chinese
Evaluated Nuclear Data Library (CENDL-3.1)", Proc. International Conference
on Nuclear Data for Science and Technology, Jeju Island, Korea, April 26-30, 2010 (in press).
\end{thebibliography}
\end{document}